\documentclass[aps,pra,showpacs,twocolumn,superscriptaddress,nolongbibliography,floatfix]{revtex4-2}
\usepackage{graphicx}
\usepackage{amsmath}
\usepackage{xparse}
\usepackage{dcolumn}
\usepackage{bm}
\usepackage[colorlinks=true, citecolor=blue,allcolors=blue]{hyperref}
\usepackage{physics}

\begin{document}
\title{A pump-probe experiment in cw-mode on ionization of Rydberg atoms
} 

\author{K.\,L.~Romans}
\affiliation{Physics Department and LAMOR, Missouri University of Science \& Technology, Rolla, MO 65409, USA}

\author{ B.\,P.~Acharya}
\affiliation{Physics Department and LAMOR, Missouri University of Science \& Technology, Rolla, MO 65409, USA}

\author{ A.\,H.\,N.\,C.~De Silva}
\affiliation{Physics Department and LAMOR, Missouri University of Science \& Technology, Rolla, MO 65409, USA}

\author{ K.~Foster}
\affiliation{Physics Department and LAMOR, Missouri University of Science \& Technology, Rolla, MO 65409, USA}

\author{ O.~Russ}
\affiliation{Physics Department and LAMOR, Missouri University of Science \& Technology, Rolla, MO 65409, USA}

\author{D.~Fischer}
\affiliation{Physics Department and LAMOR, Missouri University of Science \& Technology, Rolla, MO 65409, USA}

\date{\today}

\begin{abstract}
Rydberg atoms are in the focus of intense research due to the peculiar properties which make them interesting candidates for quantum optics and quantum information applications. In this work we study the ionization of Rydberg atoms due to their interaction with a trapping laser field. A reaction microscope (ReMi) is used to measure photoelectron angular and energy distributions. Reaction microscopes are powerful tools when brandished against atomic photoionization processes involving pulsed lasers; the timing tied to each pulse is crucial in solving the subsequent equations of motion for the atomic fragments in the spectrometer field. However, when used in pump-probe schemes, which rely on continuous wave (cw) probe lasers, vital information linked to the time of flight is lost. This study reports on a method in which the standard ReMi technique is extended to cw-mode probes through coincidence measurements. This is then applied to the photoionization of $^6$Li atoms initially prepared in optically pumped $2^{2}S_{1/2}$ and $2^{2}P_{3/2}$ states. Multi-photon excitation from a tunable femtosecond laser is exploited to produce Rydberg atoms inside an infrared optical dipole trap; the structure and dynamics of the subsequent cascade back towards ground is evaluated.
\end{abstract}


\maketitle

\section{Introduction}
Rydberg atoms, i.e., atoms with one electron populating a highly excited state, have been the subject of intense experimental and theoretical research for many years. These systems have a number of peculiar properties including comparably long lifetimes, large spatial extensions that can be many orders of magnitude greater than the size of ground state atoms, a strong response to external electric and magnetic fields, and they exhibit controllable long-range interactions. In the focus of fundamental research are, e.g., Rydberg blockade \cite{Lukin2001,Urban2009}, Rydberg--Rydberg spatial correlations \cite{Schwarzkopf2011}, many-body states \cite{Browaeys2020}, and the formation and properties of exotic Rydberg dimers \cite{Boisseau2002,Bendkowsky2009,Overstreet2009,Hollerith2019,Zuber2022}. On the applied side, Rydberg atoms are considered well-suited platforms for quantum optics \cite{Tiarks2018,Khazali2019}, quantum computation \cite{Saffman2010,Khazali2020}, and quantum simulation \cite{Weimer2010}.

Experimentally, Rydberg atoms are typically prepared by laser-cooling an atomic sample, exciting some of the atoms by a combination of resonant laser beams to highly excited states, and trapping them in optical dipole traps or in optical lattices. In these schemes, the binding energy of the Rydberg atoms can be several orders of magnitude below the photon energy of the trapping lasers resulting in photoionization. This leads to noticeable ionization rates (compared with spontaneous decay) \cite{Cardman2021} and limits the fidelity of Rydberg quantum-control schemes. Therefore, a detailed understanding of photoionization processes of Rydberg atoms in optical trapping laser fields is desirable.

In the present work, we investigated the photoionization of $^6$Li Rydberg atoms out of an optical dipole trap in a kinematically complete experiment. The lithium atoms are first cooled and trapped in a near resonant all-optical trap \cite{Sharma2018} creating a sample of cold atoms being in the 2$S$ ground state or in a polarized 2$P$ state. The atoms are subjected to femtosecond laser pulse exciting the valence electron to states with principal quantum numbers of $n
\geq 6$. During the whole process, the atoms are exposed to the continuous wave field of an infrared dipole trap laser. The fragments of atoms ionized in this field are collected and momentum-analyzed in a reaction microscope \cite{Doerner2000,Ullrich2003,Fischer2019}.

Notably, a new data-analysis algorithm was developed that allows the reconstruction of the ionization time of the atoms in the cw field with a resolution of few nanoseconds and thereby enables extracting the 3-dimensional momentum vectors of the target fragments. Therefore, not only are the photoelectron angular distributions accessible but also the time-dependent population dynamics due to transitions caused by spontaneous decay and photoionization. To our knowledge, the present study constitutes the first fully differential experiment on photoionization of Rydberg atoms and it provides a path to get new insights into the dynamics of Rydberg states and into light--matter interaction, where the wavelength of the ionizing radiation becomes comparable to the size of the target.

\section{Experimental Setup}
This experiment is comprised of four key components: A derivative of the standard magneto-optical trap (MOT) known as the all-optical trap (AOT) \cite{Sharma2018}, a pulsed femtosecond (fs) laser source in a tunable mode, a continuous wave optical dipole trap (ODT) laser source, and a reaction microscope (ReMi) capable of measuring the momentum of ionization fragments over the full solid angle.

\subsection{Magneto-optical trap} 
In order to understand the AOT a brief review of the standard MOT, which is a standard experimental technique routinely used in many laboratories, is provided below. Specific details on the present hardware can be found in \cite{Hubele2015}.

The atom trap is contained in a vacuum chamber with pressure on the order of $10^{-10}$ mbar and is fed from an atomic $^6$Li loading system similar to \cite{Tiecke2009}. In the center of the chamber are two sets of copper coils set up in the anti-Helmholtz configuration to create a quadrupole magnetic field whose magnitude grows linearly from its central zero point. Located here is the intersection of the trapping laser beams provided by a ECDL (external cavity diode laser) followed by a tapered-amplifier (TA 100 from Toptica Photonics) whose frequency is red-shifted down from the $^6$Li D$_{2}$-line, $2^2S_{1/2}$ $\rightarrow$ $2^2P_{3/2}$ ($\lambda$ $\approx$ 671 nm). The ECDL's single output is separated into three beams (for three orthogonal axes) where each is sent through a combination of polarizers, $\lambda/4$ plates, and back reflecting mirrors to create a total of six beams. The resulting three beam-pairs each have opposite helicity, ($\sigma^{+}$ - $\sigma^{-}$). 

Both the ground and excited states experience hyperfine splitting of their energy levels on the order of 228 MHz and 5 MHz respectively; this makes the primary transition the $2^2S_{1/2}$ ($F = 3/2$) $\rightarrow$ $2^2P_{3/2}$ ($F = 5/2$) channel and constitutes the "cooler" frequency. To insure that atoms are not lost to the dark state $2^2S_{1/2}$ ($F = 1/2$) a second frequency is introduced to form the "repumper". This coherent superposition of frequencies is achieved using an electro-optical phase modulator (EOM) that creates sidebands shifted about $\pm$228 MHz with respect to the central cooler frequency, with power distributed as 50\% in the cooler and 25\% in each of the two sidebands. The sideband that is blue-shifted up 228 MHz is used as the repumper while the sideband that is red-shifted down falls far enough from any resonance that it is ignored.

The main trapping mechanism in magneto-optical traps is well understood and described in the literature (see, e.g., \cite{Metcalf2007}). Briefly, as the atoms approach the intersecting trapping lasers they will begin to absorb and scatter photons. Each absorbed photon imparts momentum $\hbar \boldsymbol{k}$ to the atom in the direction of its propagation, where $\boldsymbol{k}$ is the wavevector whose magnitude is related to the wavelength by $k = 2\pi/\lambda$. This photon is then spontaneously emitted (or scattered) by the atom after a short time into a random direction. The effect of these spontaneous emissions averages out to zero over the time scale of the experiment and does not contribute to the overall momentum transfer. 

To qualitatively illustrate the main mechanism of the trap, consider a two-level atom whose excited state is associated with a scattering rate $\Gamma$, 
which is related to the state lifetime, $\tau = 1/\Gamma$. Repeated absorption over time generates an average force proportional to $\hbar \boldsymbol{k}\:\Gamma$. The proportionality constant depends on the intensity of the beam, $I$, and the detuning of the laser frequency with respect to the transition resonance, $\Delta$. If one of these atoms is moving with a velocity $\boldsymbol{v}$ along the axis of a single ($\sigma^{+}$ - $\sigma^{-}$) beam-pair, then this proportionality constant can be seen in the following equation of the force \cite{Petrov2006, Metcalf2007},
\begin{equation}
\boldsymbol{F}_{\pm} = \hbar \boldsymbol{k}_{\pm}\Gamma \left( \frac{ s_{0}/2}{1 + s_{0} + (2\Delta_{\pm}/\Gamma)} \right),
\label{eq:force}
\end{equation}
with the on-resonance saturation parameter,
\[ s_{0} \equiv I/I_{s}, \]
the saturation intensity,
\[ I_{s} \equiv \pi h c\Gamma/3\lambda^3 ,\]
and the total detuning for each beam,
\[ \Delta_{\pm} = \delta \mp \boldsymbol{k} \cdot \boldsymbol{v} \pm \frac{(\boldsymbol{\mu}_{e}-\boldsymbol{\mu}_{g}) \cdot \boldsymbol{B}(\boldsymbol{r})}{\hbar}. \]

The first term $\delta$ is the standard detuning that we control directly. The term due to the Doppler effect depends on the motion of the atom relative to the beam's propagation direction. This Doppler term establishes a velocity dependent force that dissipates energy from the atoms and cools them. Finally, a term due to the Zeeman shift of the atomic transition. Assuming that the nuclear magnetic moment can be ignored, it is approximately proportional to the dot product of the magnetic field $\boldsymbol{B}$ and the atomic magnetic moment $\boldsymbol{\mu}$ of the excited and ground states. The Zeeman term creates a restoring force that both depends on position $\boldsymbol{r}$ and the magnetic quantum number $m$. Combined, the Doppler and Zeeman terms act to simultaneously cool the atoms to the mK range (near the 0.2 mK Doppler limit of the D$_{2}$-line) and confine them to a volume on the order of 1 mm$^{3}$ resulting in atomic densities up to $10^{12}$\,atoms/cm$^{3}$.

The standard MOT is a robust tool that has had great success in trapping Alkali atoms \cite{Raab1987}, and Lithium specifically \cite{Lin1991}, to be used as targets in collision studies. Of particular interest has been  their combination with ReMi to study recoil-ion and electron momenta in ionization experiments (e.g. \cite{Zhu2009}). However, the quadrupole magnetic field frustrates measurements of the electrons' momenta and needs to be quickly switched off before ionization.
Furthermore, the electrons' motion after ionization must be constrained to the chamber via the addition of a homogeneous magnetic field (see section~\ref{sec:spec}). Along the axis of the overlapped magnetic fields the corresponding laser beam pair's polarization and central offsets must be adjusted to compensate for the shift in the magnetic field's zero point. As a result, different  magnetic sublevels are populated in the cooling cycle leading atoms to "leak" out along the adjusted axis. This modified trap has been dubbed the 2.5D MOT \cite{Hubele2015}, which yields a reduced atomic density up to $10^{9}\,$atoms/cm$^{3}$ and duty cycles upwards of 50\%. Both the standard MOT and 2.5D MOT tend to leave the target overall unpolarized, i.e., it is not easily possible (without a dedicated excitation laser beam) to prepare the atoms with an aligned angular momentum. 

\subsection{All-optical trap}
To counter the above difficulties the MOT was modified into the AOT. More details can be found in \cite{Sharma2018}
but the important differences are summarized below.

We begin with a standard MOT with the goal of completely reducing the quadrupole field to zero. This is done by a series of iterative steps wherein each one the magnitude of the quadrupole field is decreased and the polarizations, central offsets, and relative intensities of each beam-pair are adjusted to retain a signal in the atom trap. These corrections are then optimized at each step to gain the maximum signal before another reduction is 
made. Steps are taken until the quadrupole field is completely off leaving the system in one of many possible configurations in parameter space. Once established, the AOT can be changed between configurations with significantly less effort.

The AOT is highly asymmetric in beam pair parameters, especially compared to the standard MOT. On the other hand, this technique yields temperatures and atom densities comparable to the 2.5D MOT without a quadrupole magnetic field or associated switching cycle; the trapping and cooling is now done entirely with laser fields. Into the bargain, each configuration leads to a significant degree of polarization in the target that can be taken advantage of as initial state preparation. The configuration used in this study is similar to the $\sigma$-configuration used in \cite{Sharma2018}. Consequently, the atoms are optically pumped such that both the ground and excited states driven in the trapping transition are maximally aligned (i.e. $|F,m_{F}\rangle$ = $|F,F\,\rangle$\,). 

Now, $\boldsymbol{B}(\boldsymbol{r}$) is no longer position dependent and according to Eq.~\eqref{eq:force} trapping cannot be achieved by the Doppler term alone (also see \cite{Ashkin1983}). 
While the trapping mechanism is not fully understood, the Bi-chromatic Rectified Dipole Force (BRF), originally proposed in \cite{Kazantsev1987,Kazantsev1989} and later explored by \cite{Grimm1990,Yatsenko2004,Zhang2014}, is the most plausible candidate. A position dependent intensity field can arise from the interference of the cooler and repumper frequencies along the beam-pairs, which is heavily dependent on the full ensemble of beam parameters. The cooler drives the primary trapping transition while the repumper "rectifies" or modulates the cooler's amplitude in space. The corresponding energy shifts in the atomic levels due to this rectification will lead to the sought after position dependence. Lastly, the offsets for each beam as well as the imbalanced powers for each pair generates an additional dissipative "vortex force" that drives the atoms around a central axis, similar to the monochromatic version discussed in \cite{Walker1992}.  

\subsection{Femtosecond laser}
\begin{figure}[t]
\centering
\includegraphics[keepaspectratio, width=0.98\linewidth]{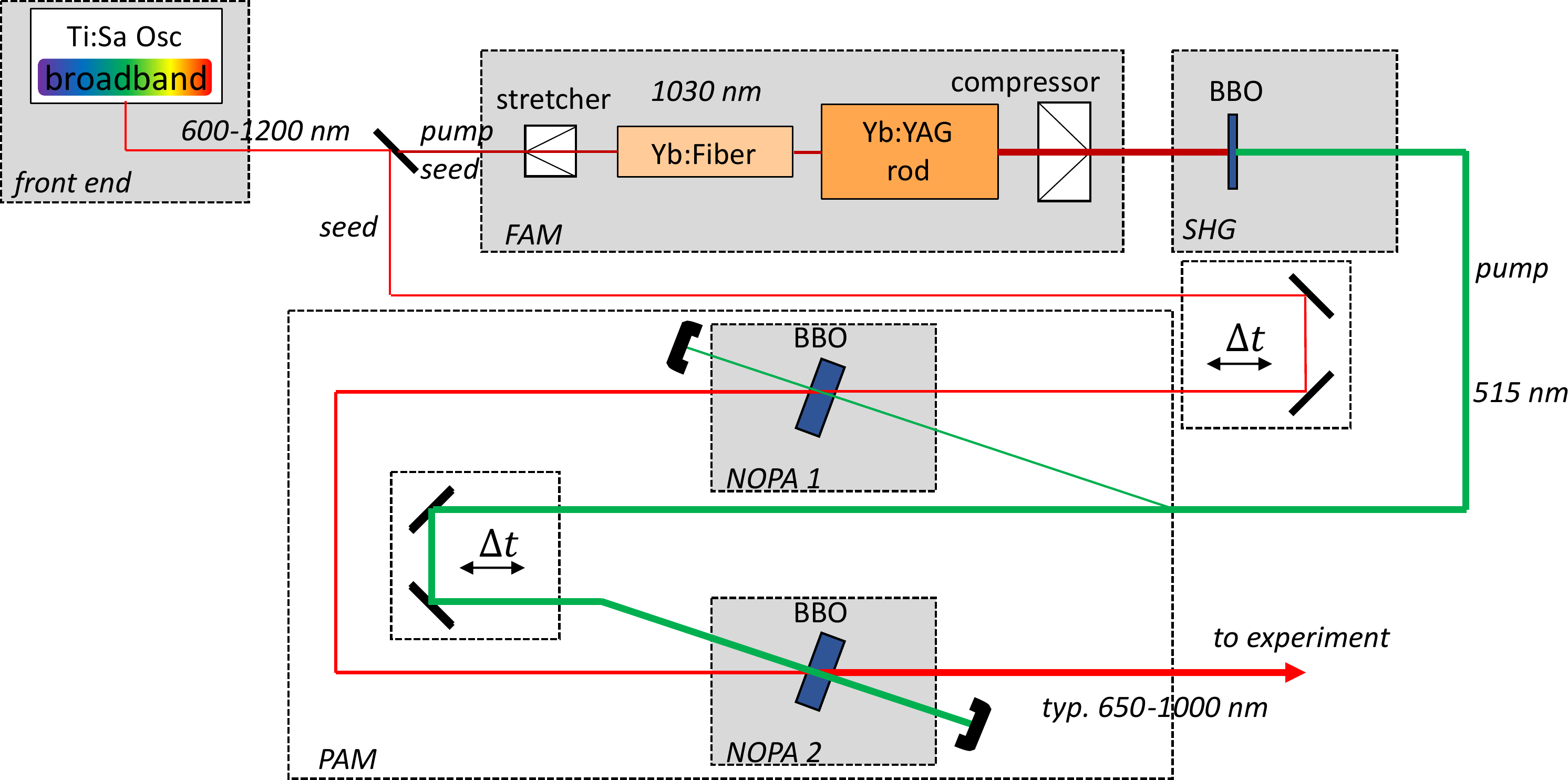}
\caption{Schematic of the OPCPA. After the oscillator, the FAM (top) creates the pump beam that the PAM (bottom) uses to amplify the seed-signal. Delay stages adjust temporal overlap $\Delta t$ between pump and seed at each NOPA crystal.}
\label{fig:OPCPA}
\end{figure}

To conduct the ionization experiments a commercially available few-cycle optical parametric chirped-pulsed amplifier (OPCPA), built by the company Laser Quantum, was wielded similar to the setup described in \cite{Harth2017}. A schematic of the system can be seen in Fig.~\ref{fig:OPCPA}.

The front end begins with a Ti:Sapphire crystal oscillator driven by a diode-pumped solid-state laser at 532 nm in a collimated fundamental Gaussian mode. The crystal is oriented at its Brewster angle to minimize reflections and to define the horizontal (parallel to the table) for polarization. At this point, the laser is running continuously and the formation of ultra-short pulses is achieved using dispersion-managed mode locking via the Kerr Effect in the crystal (e.g. \cite{Duguay1969}).

The broadband gain medium ranges from 600 - 1200 nm, yielding pulses roughly 5 fs in width, 2.5 nJ average energy, at a repetition rate of 80 MHz. This output from the oscillator then has a small infrared (IR) part filtered out ($\approx$ 1020 - 1060 nm) to be used as a pump further down the apparatus while the remainder becomes the seed-signal to be amplified. In the path of the seed-signal is an additional optical medium of quartz and fused-silica used to stretch it temporally to an approximately 1000 fs pulse for a tunable-mode technique applied later. 

The IR part is sent through a three-stage Fibre Amplifier Module (FAM) followed by a rod-type amplifier which is pumped by an additional diode laser at 976 nm. After compression, it hits a Second-Harmonic Generation (SHG) non-linear crystal yielding a 30 W average power, 200 kHz, 515 nm pulse to be used as the pump for the next module. 

The seed-signal is sent through the Parametric Amplifier Module (PAM) containing two non-collinear optical parametric amplifying (NOPA) crystals of Beta Barium Borate (BBO). In each NOPA crystal the stretched seed-signal is spatially and temporally overlapped with part of the pump created in the FAM. Within this overlap energy from the pump beam is transferred to the seed while conserving total energy, thus making the process parametric. The two beams must also intersect at a "magic angle" within the crystal to maximize the effect at phase-matching conditions. Due to energy and momentum conservation, a third "idler" beam is generated in the process (which is dumped and not used). In our configuration, the seed beam is chirped and stretched in time such that the pump is much shorter in width. Therefore, only that part of the seed spectrum contained in the overlap is amplified. In this way we can tune the central-peak wavelength that is amplified in the crystal by adjusting the overlap.

After the seed-signal passes through both NOPA stages it outputs with an average power up to 3 W, a central-peak wavelength tunable between 660 - 1000 nm (740 $\pm$ 20 nm for this experiment), and has a pulse roughly 150 fs in width. This output is then compressed by chirped mirrors before being sent to the experiment chamber. In the present experiment, the beam is circularly polarized (co-rotating with the electron current density of the excited atomic $2^2P_{3/2}$ state in the AOT) and focused to yield a waist of about 50 $\mu$m and average power of 150 mW at the reaction volume. 

\subsection{Optical dipole trap laser}
The ODT laser source is an industrial grade IGP Photonics YLR-series fiber laser. It is a diode-pumped Ytterbium fiber laser with a continuous wave output, a power range from 20 - 200 W, and a wavelength of 1070 $\pm$ 5 nm. During the experiment this laser was run at 100 W and overlapped with the fs-laser in the reaction volume with a focus waist that was also about 50 $\mu$m. This laser generates an ODT inside the AOT that modifies the environment the atoms are in and thereby opens up new ionization channels to them. 

\subsection{Momentum spectrometer}
\label{sec:spec}
More detail on the spectrometer can be found in \cite{Hubele2015} and on the ReMi in general in \cite{Moshammer2003,Fischer2019}. An outline is given below. 

The spectrometer consists of three primary components: First, are the coils outside the chamber in the Helmholtz configuration that are used in the collection of electrons. The symmetry axis of these coils defines the longitudinal, or z-axis, as well as the quantization axis for the atoms. Next, ring electrodes run the length of the chamber that facilitate the generation of electric fields in a variety of arrangements and geometries (a uniform longitudinal field  was used here). Finally, at each end of the spectrometer are the position-sensitive detectors formed from microchannel plates (MCP) combined with delay-line anodes whose face defines the transverse or xy-plane. 

These detectors not only yield the xy-position for each ionization fragment, ($x_{d}, y_{d}$), but the time of flight as well which is defined by the time of ionization and the time for each fragment to reach the MCP,
\begin{equation}
    T = t_{mcp} - t_{ion}    
    \label{eq:time_of_flight}.
\end{equation}
The ionization time is associated with the measurement of the fs pulse striking a photodiode ($t_{ph}$) outside the chamber. In the following discussion, we simply ignore differences in these times due to the light travel time from the diode to the experiment chamber and due to cable delays, which simply result in constant offsets (i.e., $t_{ion}=t_{ph}$). Lastly, the fixed distance from each detector to the center of the experiment chamber defines the distance $z_{d}$.

Since both homogeneous electric and magnetic fields are overlapped along the longitudinal axis each ionization fragment will be subject to the well known equations of cyclotron motion. These equations can be solved for the initial momenta components to yield (see also \cite{Fischer2019}):
\begin{subequations}
     \begin{equation}
         p_{0x} = \frac{ q B_{z} }{2} ( \cot(\frac{\omega_{c} T}{2})\,x_{d} - y_{d} ), 
     \end{equation}
      \begin{equation}
          p_{0y} = \frac{ q B_{z} }{2} ( x_{d} + \cot(\frac{\omega_{c} T}{2})\, y_{d} ), 
      \end{equation}
      \begin{equation}
          p_{0z} = z_{d}\, \frac{m}{T} - \frac{q E_{z}}{2}\,T. \hspace{4em}\label{eq:zmomentum}
      \end{equation}  
      \label{eq:cyclotron_motion}
\end{subequations}\\
Here, $\omega_{c} = q B_{z}/m_{q}$ is the cyclotron frequency for a charge $q$ with mass $m_{q}$ that is subject to both longitudinal fields, $B_{z}$ and $E_{z}$. It can be readily seen that with the detector coordinates $(x_{d},y_{d},z_{d})$ and the time of flight $T$, all three initial momentum components for each fragment are uniquely determined. On a final note, the electric field magnitude was chosen such that each electron arrives at the detector within a single cyclotron period, $T_{c} = 2\pi/\omega_{c}$, from ionization. 

\section{Data Analysis}
A new challenge arises once the pulsed fs and continuous ODT laser fields are overlapped as ionization from the ODT can happen at any time between fs pulses. Ergo, the time of ionization $t_{ion}$ is now unknown and can no longer be associated with the photodiode measurement ($t_{ion} \neq t_{ph}$). However, the Coincidence measurements of the photoelectron and recoil-ion, along with momentum conservation, will help solve this obstacle. 

Between the fs-pulses is a 5$\,\mu$s window in which an ensemble of data ($x_{d}, y_{d}, t_{mcp}$) is gathered for each recoil-ion and electron. While the exact time of flight $T$ of each individual particle is not directly measured, the range of $T$ is known and it depends essentially only on the spectrometer settings and the range of observed photo-electron $z$-momenta (cf.\ Eq.~(\ref{eq:zmomentum})). Using Eq.~(\ref{eq:time_of_flight}), the range of possible ionization times $t_{ion}$ is calculated. In a first step, $t_{ion}$ is fixed to the smallest value in that range and used with Eq.~(\ref{eq:cyclotron_motion}) to calculate the values of the recoil ion momentum, $\boldsymbol{p}_r$, and its coincidentally measured electron momentum, $\boldsymbol{p}_{e}$. The total momentum $\boldsymbol{Q}$ is thus,
\begin{equation}
    \boldsymbol{Q} = \boldsymbol{p}_{r} + \boldsymbol{p}_{e}. 
\end{equation}
Due to momentum conservation, $\boldsymbol{Q}$ is equal to the ionizing photon's momentum, which is negligible. Therefore, the parameter $t_{ion}$ is varied until $\boldsymbol{Q}$ is minimized for that event. This scheme is then applied to all events in the set to recover the unknown ionization times. Ideally, $\boldsymbol{Q}$ is the zero. In practice however, $\boldsymbol{Q}$ retains a finite minimized value that contributes to the upper bound on the time resolution.

In essence, this method relies on the fact the the masses of the electron and the recoil ion are very different (by about four orders of magnitude). As a result, the arguments of the cotangents in Eq.~(\ref{eq:cyclotron_motion}), $\omega_{c} T/2$, which describe the spiral cyclotron motion of the particles in the magnetic spectrometer field, span a very different angular range for electrons and recoil ions. In our experiments, the magnetic field strength $B_z$ is chosen such that the electron cyclotron phase defined as $\omega_{c} T$ covers a range of nearly $2\pi$. For the ions, this range is much smaller and the cyclotron phase angle $\omega_{c} T$ varies only insignificantly. Therefore, the relative positions of electrons and recoil ions (in particular, their relative azimuthal angles) provide information on the cyclotron phase of the electron, and, hence, on its time of flight $T$.  

\begin{figure}[htbp]
\centering
\includegraphics[keepaspectratio, width=0.98\linewidth]{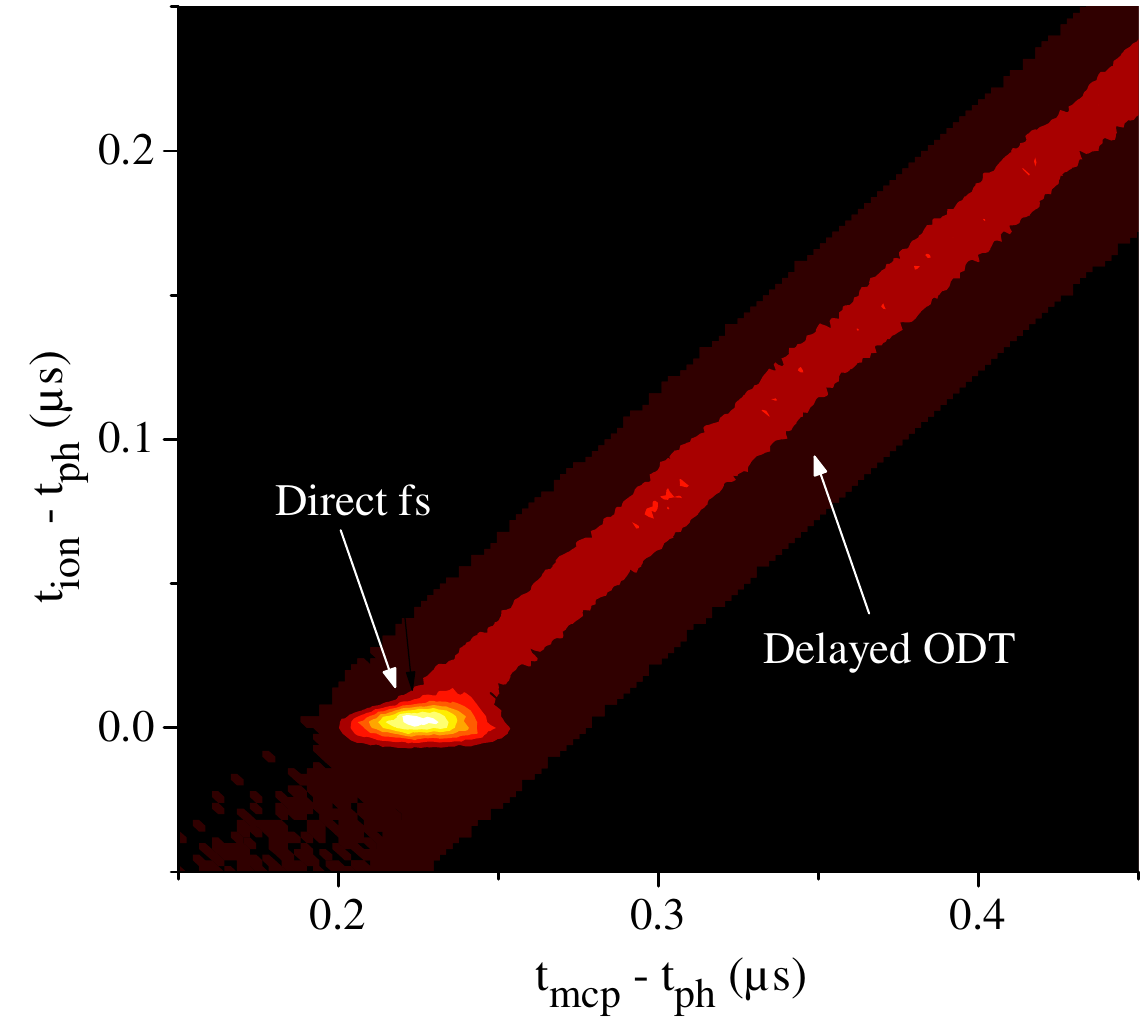} 
\caption{Derived time of ionization vs measured time of arrival with respect to photodiode. The z-axis of the plot corresponds the number of events on a linear scale.}
\label{fig:derive_vs_measured_time}
\end{figure}

As a test of this method the derived time of ionization and the measured time of arrival, both with respect to the photodiode, can be plotted against one another to give Fig.~\ref{fig:derive_vs_measured_time}. When plotted in this way it becomes clear that the above distribution mirrors Eq.~(\ref{eq:time_of_flight}). The bright spot is due to the direct fs ionization and its vertical width gives an estimate on the time resolution ($\pm$5$\,$ns), which is essentially limited by the position resolution for the recoil ions. The tail comes from the delayed ODT ionization and has a slope of unity.

\begin{figure}[htbp]
\centering
\includegraphics[keepaspectratio, width=1\linewidth]{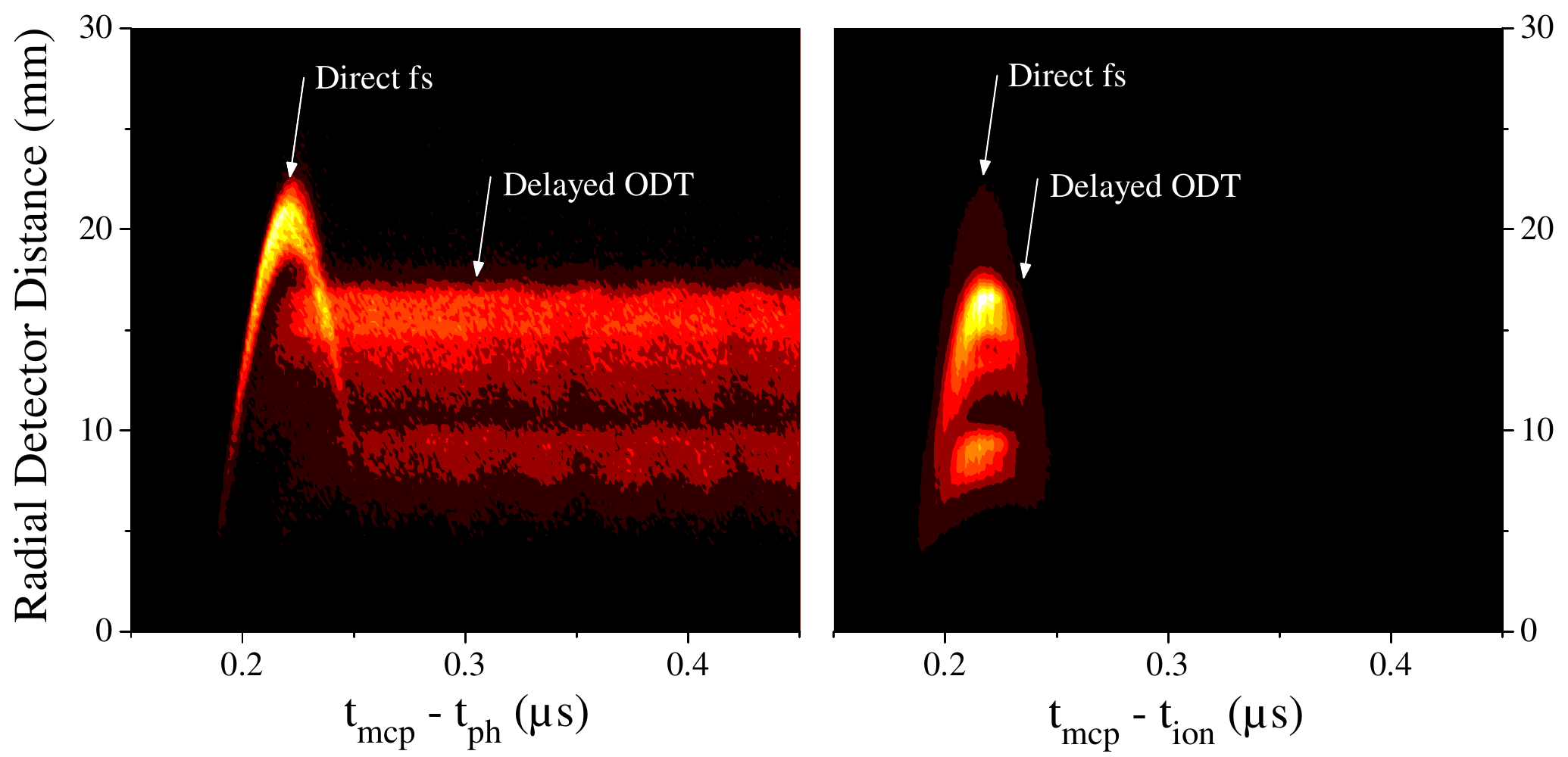} 
\caption{Electron radial detector distance vs measured time of arrival (left) and derived time of flight (right). The z-axis of the plot corresponds the number of events on a linear scale.}
\label{fig:radial_vs_time}
\end{figure}

Another check of the method can be seen by plotting the radial distance of the electrons on the detector first versus the measured time of arrival and then versus the derived time of flight (as seen in Fig.~\ref{fig:radial_vs_time}). In the first case, the bright arch comes from the direct fs ionization, while the trailing lines are due to the delayed ODT ionization. The second case highlights that the time of flight for all the fragments fall within a small time window set by the fixed spectrometer properties and photoelectron energy range. These plots suggest that the fs-pulse first excites the atom to some state(s) and then the ODT ionizes the states participating in the resulting cascade. The internal structure of the atom also becomes evident as the radial bands, seen on the left of Fig.~\ref{fig:radial_vs_time}, emerge from different atomic energy levels. 

\section{Results and Discussion}
Now that the time of flight can be recovered for any event, the ionization process can be further illuminated.

\begin{figure}[htbp]
\centering
\includegraphics[keepaspectratio, width=1\linewidth]{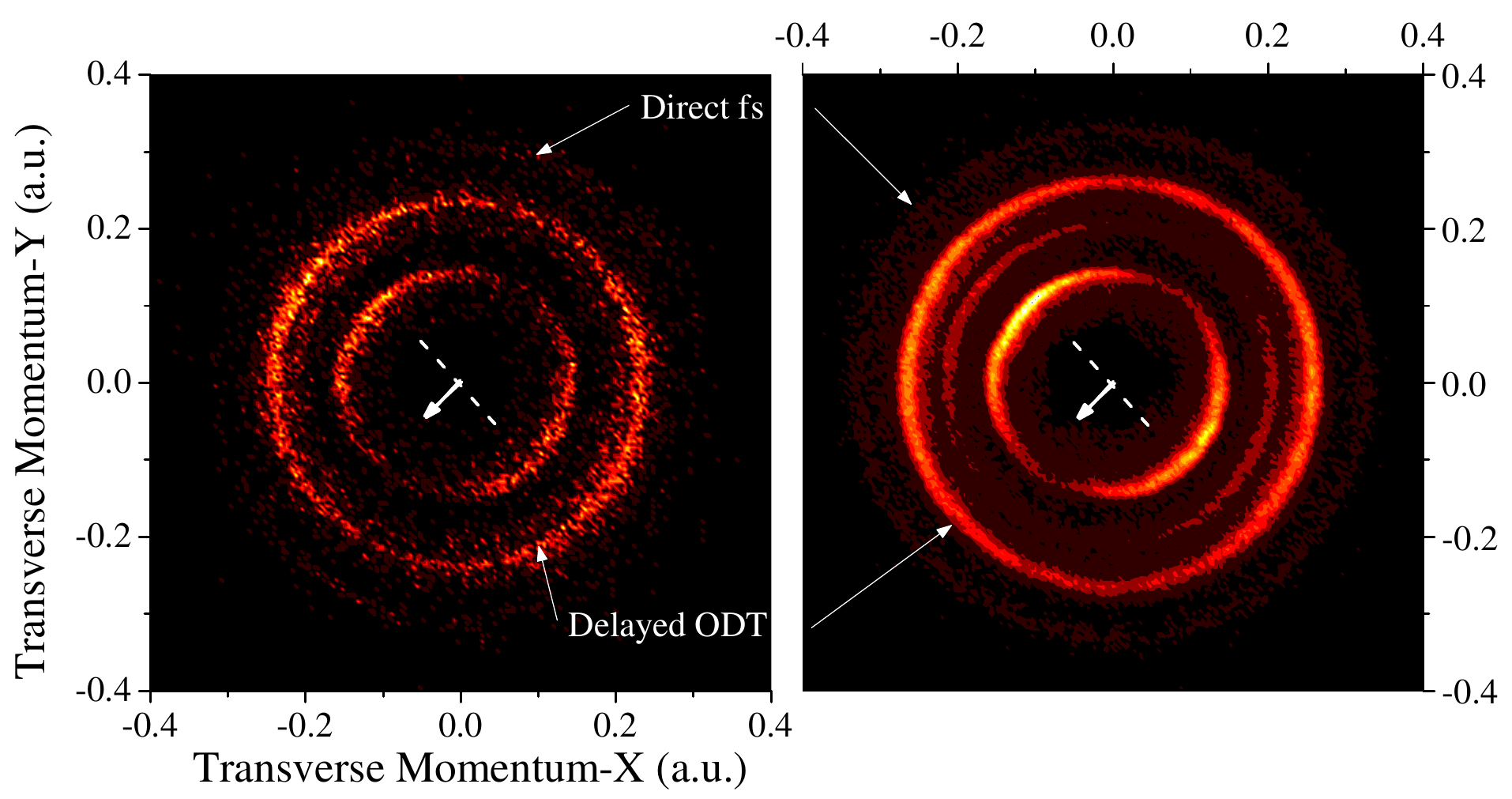} 
    \caption{Electron transverse momentum distributions using derived times from the $2S$ (left) and $2P$ (right) initial states, in atomic units. ODT propagation direction (arrow) and linear polarization axis (dashes) are denoted at each center. The fs-pulse is propagating along the axis into the page. The z-axis of the plot corresponds the number of events on a linear scale.}
\label{fig:transverse_momentum}
\end{figure}

With the derived time of flight Eq.~(\ref{eq:cyclotron_motion}) can be applied again to reconstruct the electron momentum spectra. In Fig.~\ref{fig:transverse_momentum}, the transverse momentum for the co-rotating fs pulse with the linear ODT is shown for  both the $2^{2}S_{1/2}$ and $2^{2}P_{3/2}$ initial states (hereafter referred to as $2S$ and $2P$). The radius of each ring corresponds to a given photoelectron energy and the distribution of the ring about the xy-plane conveys information on the final angular state(s). 

\begin{figure}[htbp]
\centering
\includegraphics[keepaspectratio, width=0.98\linewidth]{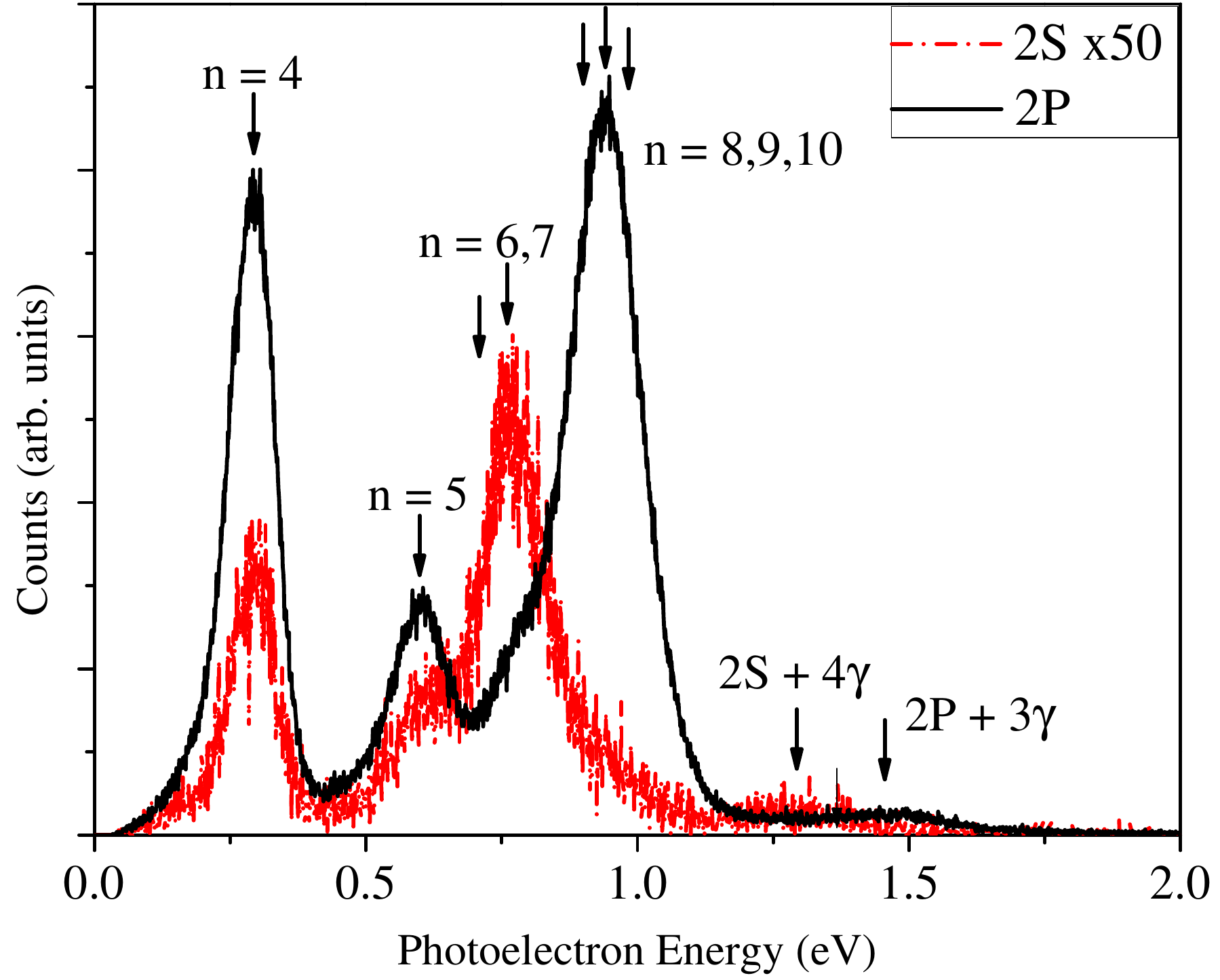} 
\caption{Photo-electron energy distributions for the 2$S$ (red line) and the 2$P$ (black line) initial states. From right-to-left, the bumps are direct fs-ionization with either three or four photons ($\gamma$), and the remaining peaks are the delayed ionization. Estimated principle quantum numbers are denoted for delayed peaks.}
\label{fig:photoelectron_energySpect}
\end{figure}

In Fig.~\ref{fig:photoelectron_energySpect}, detected counts are plotted against the photo-electron energy for both the initial states. Going from right-to-left, the small bumps at about 1.3 and 1.5\,eV correspond to the direct multi-photon ionization from the fs pulse using three or four photons ($\gamma$), respectively.  Each peak thereafter is tied to a delayed-ring on the momentum spectra and was analyzed assuming two or three fs-photons to first excite from the initial states and then a single ODT-photon to ionize.
	 
\subsection{Angular distribution}
\begin{figure}[!htbp]
    \centering
    \includegraphics[keepaspectratio, width=.70\linewidth]{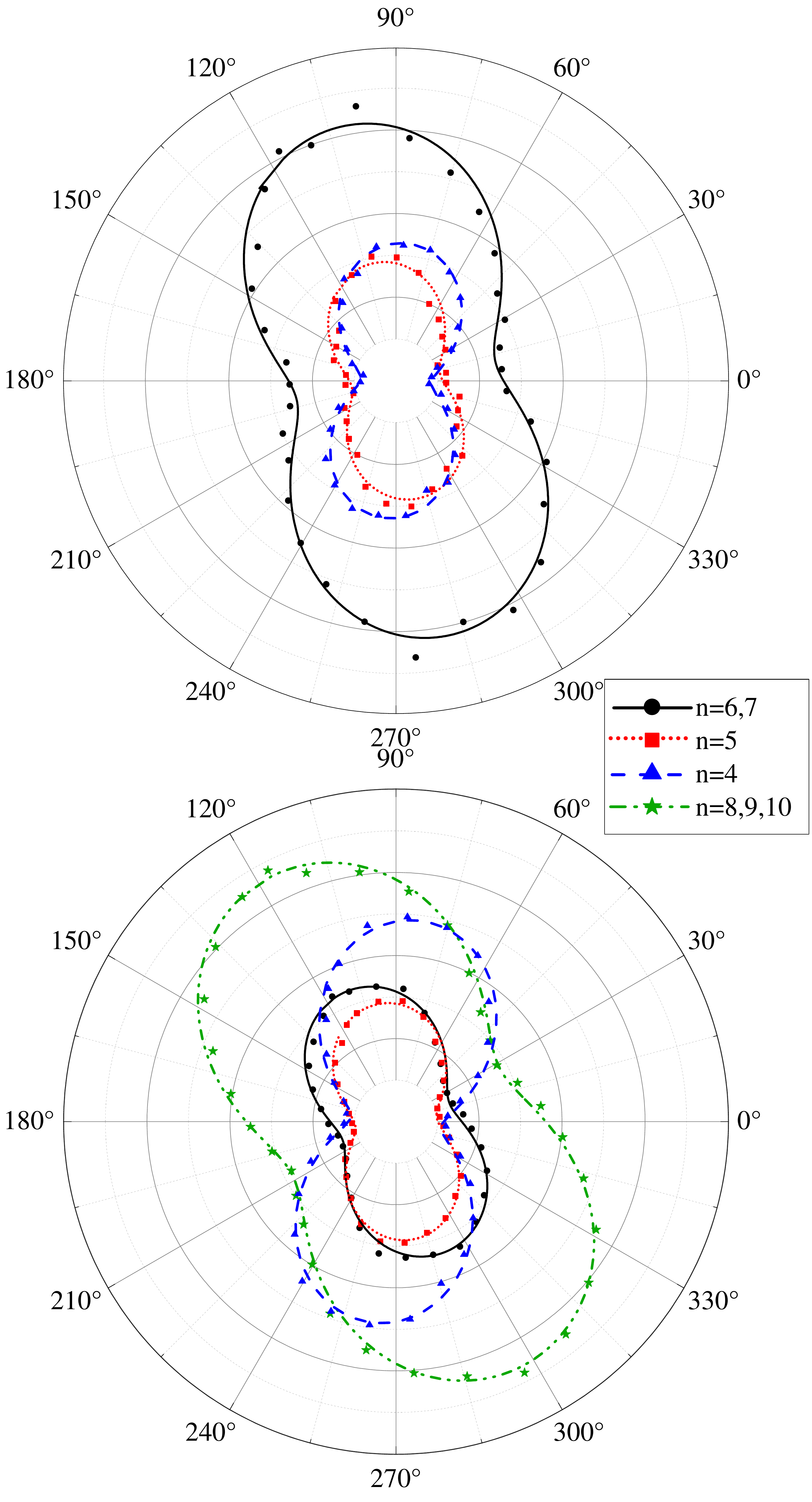}
    \caption{Photo-electron angular distributions of the delayed ionization peaks from the 2S (top) and 2P (bottom) initial states. ODT propagates radially inward along 0$^{o}$ line with polarization along 90$^{o}$ axis, in the transverse xy-plane. Quantization axis is along longitudinal z-axis into the page. Note that data points have been symmeterized.}
    \label{fig:polarplot}
\end{figure}

Each ring can first be analyzed considering the time-averaged or time-independent structure. The photo-electron angular distribution (PAD) in the transverse plane for the delayed ionization is shown in Fig.~\ref{fig:polarplot}.

For these plots, the ODT laser is propagating in the transverse plane inward along the 0$^{o}$ horizontal while its linear polarization is along the vertical axis. The quantization axis is chosen perpendicular to the plane of the plot. It is immediately clear that the PAD are rotated by an angle about the quantization axis, i.e. their major axes are misaligned to the polarization direction. This deviation can be explained by magnetic dichroism \cite{Acharya2021}. Given that the system begins optically pumped and is excited by a co-rotating laser, it is convenient to move from the ($F$, $m_{F}$) framework of hyperfine splitting into the ($\ell$, $m$) framework of the valence electron. Now, the analysis employed in \cite{Acharya2021} can be directly utilized. 

Each $n\ell$ state of interest can be ionized by the linear ODT laser into at least two final states, corresponding to both the quantum numbers, $\ell$ and $m$, changing by $\pm$1. The final state can be modelled by a partial wave expansion using
the following sum, 
\begin{equation}
\begin{split}
    \Psi_{f}(\epsilon, \theta, \phi) &= \sum_{m}\left(\sum_{l}a_{l,m}(\epsilon, \theta)\right)\, e^{i m \phi}\\
    &= \sum_{m} c_{m}(\epsilon, \theta)\, e^{i m \phi},
    \end{split}
\end{equation}
where the $c_{m}$ are the complex partial wave amplitudes, $\epsilon$ is the continuum state energy, $\theta$ and $\phi$ are the polar and azimuthal angles respectively, and the sum is taken over the allowed magnetic sublevels. For the data seen so far, the polar angle $\theta$ is fixed to 90$^o$ and $\phi$ is in the transverse plane. The PADs are then gained by the absolute value squared of this final state function,
\begin{equation}
\begin{split}    
|\Psi_{f}|^2 &= |\sum_{m} c_{m}\, e^{i m \phi}|^2 \\
&=(\sum_{m} c_{m}\, e^{i m \phi})\,(\sum_{m'} c_{m'}^{*}\, e^{-im'\phi})\\ 
&=\sum_{m} (C_{m})^2\\
&+\sum_{m<m'} 2\,C_{m}\,C_{m'}\,cos[(m-m')\phi + \varphi_{m,m'}],
\end{split}
\label{eq:pad}
\end{equation}
where the last equality is obtained by expressing the complex amplitudes in polar form, $c_{m} = C_{m}\,e^{i\varphi_{m}}$, using the real parameters $C_{m}$ and $\varphi_{m}$.
A clear $\phi$ dependent term arises in Eq.~(\ref{eq:pad}) due to the interference of the paths with differing $m$ and this determines the number of maxima in the distribution. An additional angular shift in the distribution appears due to the relative phase difference between the state amplitudes, $\varphi_{m,m'} = \varphi_{m}-\varphi_{m'}$, which characterizes the phenomenon of magnetic dichroism. This term rotates the whole distribution in the transverse plane about the quantization axis.

In this study, assuming maximal alignment is held (i.e $|\ell,m\rangle = |\ell,\ell\rangle$), each state is ionized by a single ODT photon corresponding to $(m- m')\phi = 2\,\phi$ and two terms in the sum. Each state can then be fit to the following simplified form,
\begin{equation}
    |\Psi_{f}|^{2}_{fit} = (C_{1})^{2} + (C_{2})^{2} + 2\,C_{1}\,C_{2}\,cos[2\,\phi+C_{12}],
\end{equation}
where ($C_{1},C_{2},C_{12}$) are real-valued fitting parameters. These fits are also shown in Fig.~\ref{fig:polarplot} as the various lines and exhibit excellent agreement with the data. 

In principle, one could find the ratio of $C_{1}$ to $C_{2}$ from the ratio of the lengths associated with the major and minor axes for each distribution; this ratio relates the absolute values of the complex amplitudes. The shift of the major axis of the distribution from the vertical axis (polarization) is $C_{12}$ and  yields a direct measure of the magnetic dichroism. It must be cautioned that the assumption of a single state per ionization peak quickly breaks down (apparent in Fig.~\ref{fig:photoelectron_energySpect}) as the highest energy peak is a superposition of closely spaced Rydberg states and the lowest is most likely an admixture of states from the resulting cascade. For such mixed systems more work would need to be done to isolate each state. 

\subsection{Time dependent population dynamics}
With the derived ionization times at hand for each fs-pulse, the time evolution of the state populations, which is governed by spontaneous decay and by photoionization of the excited states, can be inspected. In what follows, the dressing of the atomic energy levels, due to the presence of the oscillating fs and ODT fields, is neglected.

\begin{figure}[htbp]
    \centering
    \includegraphics[keepaspectratio, width=1.0\linewidth]{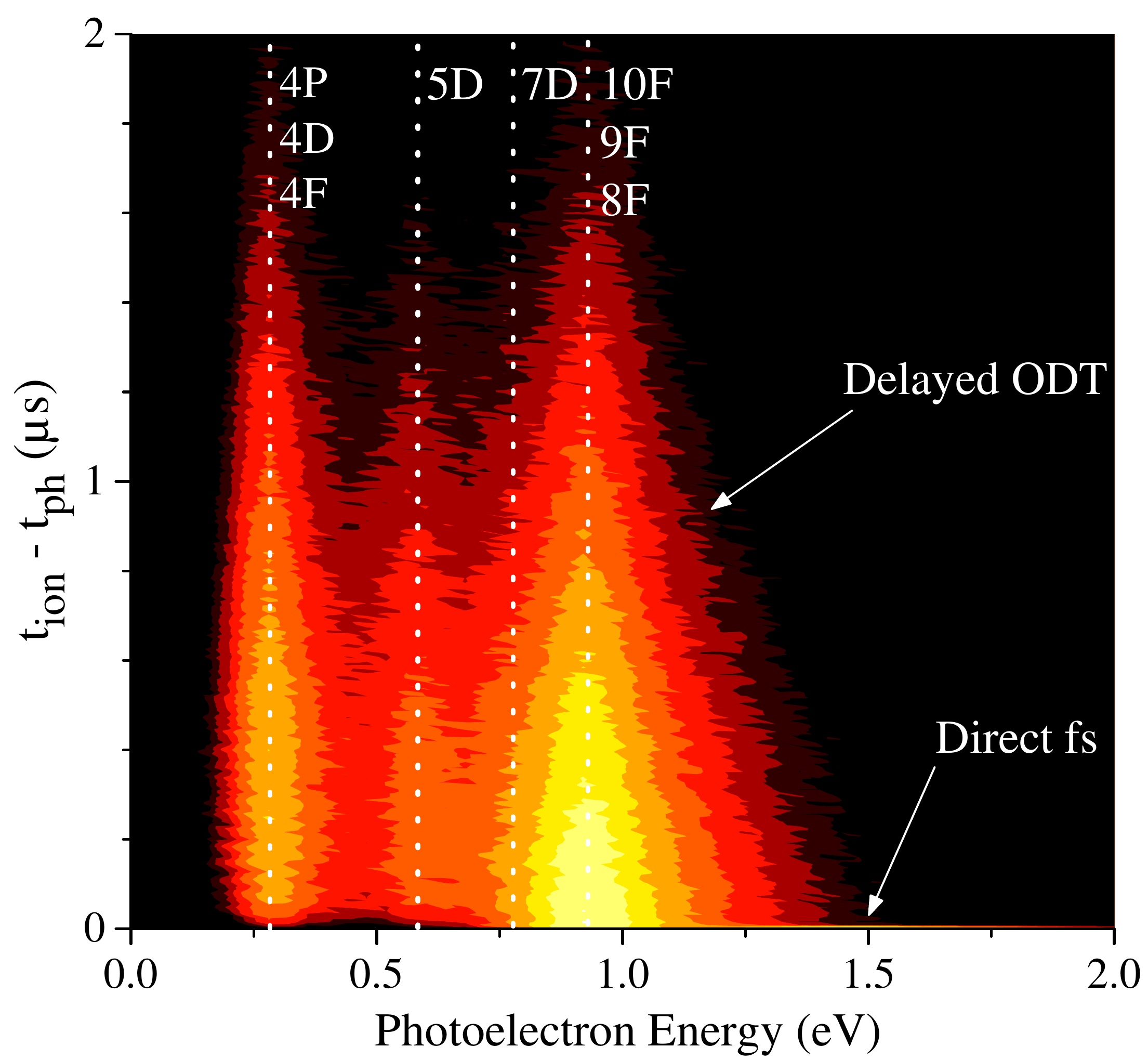}
    \caption{Derived time of ionization vs. photoelectron energy. Centers of delayed energy peaks denoted by dashed liens. The z-axis corresponds to the ionization rate and is on a logarithmic scale.}
    \label{fig:timeVenergy}
\end{figure}

Ignoring the direct fs-ionization each process yields three or four peaks of interest for the $2S$ and $2P$ initial states, respectively. The co-rotating fs-photons can only drive the optically pumped atoms to other maximally aligned states. In both the $2S$ and $2P$ cases the absorption of three and two fs-photons, respectively, drive the electron to an $F$-orbital ($\ell = 3$) that then decays in time. This time dependence can be seen for the $2P$ set in Fig.~\ref{fig:timeVenergy} wherein a set of possible participating states are highlighted. These dynamics can be tackled with basic rate equations. 

Each state (denoted by the index $i$) in the system has a population of atoms, $N_{i}(t)$, whose rate of change is governed by an equation of the form: 
\begin{equation}
    \dv{t} N_{i}(t) = \sum_{j > i} \Gamma_{i,j} N_{j}(t) - (\Gamma_{i} + R_{i}) N_{i}(t),
    \label{eq:population}
\end{equation}
where the $\Gamma_{i,j}$ are the coupling constants that drive a higher energy state $j$ to decay into $i$ via a single photon spontaneous decay. The last term contains the overall decay of the $i$-th state due to its finite lifetime $\tau_{i} (= 1/\Gamma_{i})$ and photo-ionization rate $R_{i}$. Given a system of $\mathcal{N}$ connected states, there are $\mathcal{N}$ coupled first-ordered equations of the form Eq.~(\ref{eq:population}) that govern the time evolution of the system from some given initial conditions. 

This equation can be recast into a more useful form by the use of matrices. The populations $N_{i}(t)$ can be represented by a $\mathcal{N}$ component population vector $\boldsymbol{N}(t)$, the coupling constants form the off-diagonal elements of an $\mathcal{N}\cross\mathcal{N}$ interaction matrix, $\underline{\boldsymbol{G}}$, while the lifetime and photo-ionization terms form the diagonal elements such that, 
\begin{equation}
    \dv{t} \boldsymbol{N}(t) = \underline{\boldsymbol{G}} \cdot \boldsymbol{N}(t)
    \label{eq:vector_rate_equation},
\end{equation}
where 
\begin{equation}
    \boldsymbol{N}(t) = \mqty(N_{1}(t) \\ \vdots \\ N_{i}(t) \\ \vdots),
\end{equation}
and 
\begin{equation}
   \underline{\boldsymbol{G}} = \mqty(-(\Gamma_{1} + R_{1}) & \ldots & 0 & \ldots\\  \vdots & \ddots & \ldots & \ldots \\
    \Gamma_{i,1} & \ldots & -(\Gamma_{i} + R_{i}) & \ldots \\ \vdots & \ldots & \ldots & \ddots ). 
\end{equation}
Therefore, all elements above the main diagonal are zero. Note that the states are ordered by energy with the highest excited state given the $i = 1$ index.

During this decay, the system is constantly being probed via the ODT laser and the ionization rates are measured rather than the populations directly. In order to model the system with this in mind, Eq.~(\ref{eq:vector_rate_equation}) 
 must be modified. A rate matrix $\underline{\boldsymbol{R}}$ can be formed by assigning to the main diagonal the photo-ionization rates for each state,
\begin{equation}
  \underline{\boldsymbol{R}} = \mqty(R_{1} & \ldots & 0 & \ldots  \\
    \vdots &  \ddots & \ldots & \ldots   \\
    0 & \ldots & R_{i} & \ldots   \\
    \vdots & \ldots & \ldots & \ddots  ),
    \label{eq:rate_Mat}
\end{equation}
such that the ionization rate vector is $\boldsymbol{I}(t) = \underline{\boldsymbol{R}}\cdot\boldsymbol{N}(t)$. After multiplying both sides of Eq.~(\ref{eq:vector_rate_equation}) by $\underline{\boldsymbol{R}}$ it can be shown that the ionization rate equation takes the following form,
\begin{equation}
\begin{split}
    \dv{t} \boldsymbol{I}(t) &= (\underline{\boldsymbol{R}}\cdot \underline{\boldsymbol{G}})\cdot\boldsymbol{N}(t)\\    
   &= \underline{\boldsymbol{G}}\cdot\boldsymbol{I}(t) + \comm\Big{\underline{\boldsymbol{R}}}{\underline{\boldsymbol{G}}}\cdot\boldsymbol{N}(t).
    \label{eq:ionization_rate}
\end{split}    
\end{equation}
It is clear that Eq.~(\ref{eq:vector_rate_equation}) and Eq.~(\ref{eq:ionization_rate}) have nearly identical forms, but the ionization rate equation has an extra term that includes the commutator of the rate and interaction matrices. This extra term is tied directly to the population and exclusively mixes states as the commutator only contains nonzero elements below the main diagonal. The solutions to Eq.~(\ref{eq:ionization_rate}), along with Eq.~(\ref{eq:rate_Mat}), yield both the $I_{i}(t)$ and $N_{i}(t)$ components for each state contemplated in the model. 

\begin{figure}[htbp]
\centering
\includegraphics[keepaspectratio, width=1\linewidth]{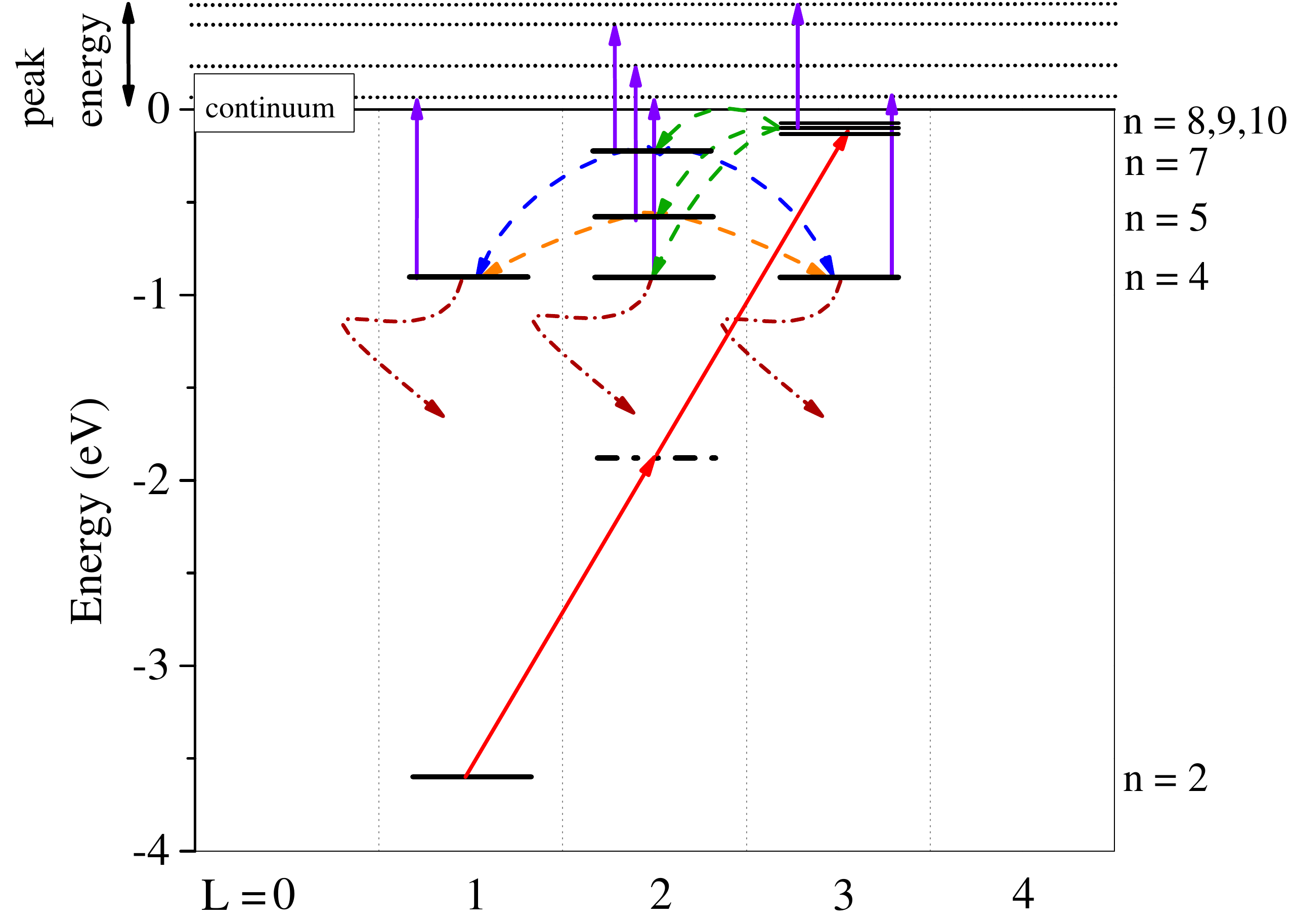} 
\caption{Principle path for $2P$ initial state. Horizontal solid/dashed black lines represent real/virtual atomic states. Other solid lines in color are transitions by the fs (diagonal, red) or ODT (vertical, purple) lasers. Curved dashed lines in color are the various spontaneous decays. Energy peaks seen in Fig.~\ref{fig:photoelectron_energySpect} are the dotted lines above the continuum threshold.}
\label{fig:2P_path}
\end{figure}

Consider the $2P$ initial state first. The fs-pulse gives two photons to the atom which send it into some $nF$-orbital. Given the width of the fs-pulse and how close large $n$ Rydberg states are in energy, it is expected that multiple states are populated. Using the center and breadth of the first peak in the $2P$ data of Fig.~\ref{fig:photoelectron_energySpect} the $n = 8,9,10$ $F$-states are chosen each containing some fraction, $f$, of the total initial population (i.e. $f_{8F}+f_{9F}+f_{10F}=1$). Continuing from right-to-left, the second peak is on the shoulder of the first (directly below the $n=6,7$ $2S$ peak) associated with the $7D$ state, the third peak with $5D$, and the fourth peak is taken as an admixture of the $4P$,$ 4D$, and $4F$ states.  This principle path is shown in Fig.\ref{fig:2P_path}

\begin{figure}[htbp]
\centering
\includegraphics[keepaspectratio, width=1\linewidth]{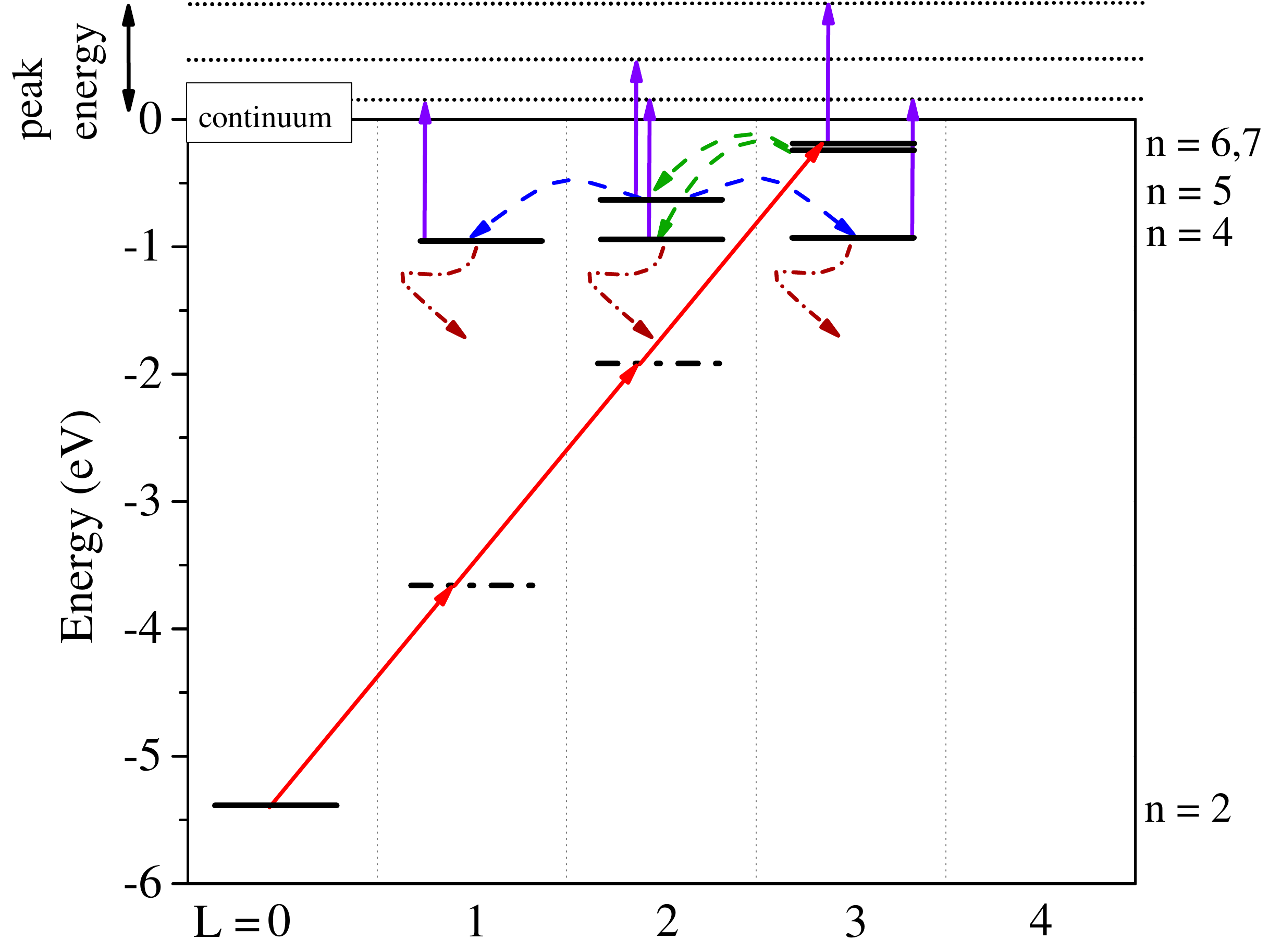} 
\caption{Principle path for $2S$ initial state. The same scheme dontes the states, decays, and transitions as Fig.~\ref{fig:2P_path}.}
\label{fig:2S_path}
\end{figure}

For the $2S$ initial case the system is excited to a mix of the n $= 6,7$ $F$-orbitals via three fs-photons which constitutes the first peak. The second $2S$ peak is tied to the $5D$ state, and the third peak is the same admixture of $n = 4$ states as the $2P$ set. The principle path for $2S$ is shown in Fig.~\ref{fig:2S_path}. Comparing both paths helps to demonstrate why the photoelectron energy spectra share significant overlap between the $2S$ and $2P$ data sets; nearly the same intermediate states are passed through as the atoms cascade back down towards the ground state. 

In order to fit the model to the data, some additional calculations and assumptions must be made. Each coupling constant $\Gamma_{i,j}$ and state lifetime $\tau_{i} (= 1/\Gamma_{i}$) for the states considered were calculated using the "Alkali.ne Rydberg Calculator" \cite{Sibalic2017}, which is a package of Python routines freely available online. The ionization rates $R_{i}$ and their ratios were calculated within the framework of a central-potential model with each atom being ionized from a single  $n\ell$-state by one photon. This begins with the following equation for the ionization cross-section found in \cite{Starace1982},
\begin{equation}
\begin{split}  
\sigma_{n \ell} &= \frac{4 \pi^2 \alpha a_{0}^2}{3} \:\frac{h\,\nu}{2\,\ell+1}\\
&\quad \cross\left[\,\ell\,|\mathcal{R}_{\ell,\ell-1}(\epsilon)|^2 + (\ell+1)\,|\mathcal{R}_{\ell,\ell+1}(\epsilon)|^2\,\right],
\end{split}
\end{equation}
where $\alpha$ is the fine structure constant ($\approx\frac{1}{137}$), $a_{0}$ is the Bohr radius ($\approx 0.529 \cdot 10^{-9}\,cm$), $h\nu$ is the ionizing photon's energy, and $\epsilon$ the photoelectron's continuum energy with the energy scale set in rydbergs ($1\,Ry = 0.5\,a.u.$). 

The radial overlap matrix elements are given by, 
\begin{equation}
    \mathcal{R}_{\ell,\ell\pm1} = \int_{0}^{\infty} \dd{r}\, U_{n,\ell}(r)\cdot r \cdot U_{\epsilon,\ell\pm1}(r),
\end{equation}
with each reduced radial function $U(r)$ given by $r$ times the full radial function of the initial discrete and final continuum state wave functions, normalized to unity and $\delta(\epsilon-\epsilon')$, respectively. The reduced radial functions were found by solving the time-independent Shcr{\"o}dinger equation numerically using the Lithium potential in \cite{Marinescu1994}, which accounts for the core polarization and quantum defect experienced by the outer valence electron,
and by a fifth-order Dormand-Prince (RKDP) method implemented in Mathematica 12.0. Finally, the photo-ionization rates for the $i^{th}$ state were obtained by multiplying the cross-section by the incoming ODT photon flux, 
\begin{equation}
    R_{i} = \Phi_{ODT}\cdot\sigma_{n_{i}l_{i}}.
\end{equation}

Each principle path solution $I_{i}(t)$ of the model takes the deceptively simple analytic form,
\begin{equation}
    I_{i}(t) = A_{i}\,e^{-(\Gamma_i + R_{i})\,t} + \sum_{j>i} A_{i,j}~e^{-(\Gamma_{j}+R_{j})\,t},
\label{eq:analytic_solution}
\end{equation}
where the amplitudes $A_{i}$ and $A_{i,j}$ are cumbersome combinations of the various $\Gamma_{i}$, $\Gamma_{i,j}$, and $R_{i}$. The ratio $A_{i,j}/A_{i}$ itself corresponds to the strength of a decay channel $j$ feeding into the state $i$ relative to the total decay channel out of $i$. For the highest excited states all the $A_{i,j}=0$, and the $A_{i}$ reduce to $I_{i}(0)\,f_{i} = R_{i}\,N_{i}(0)\,f_{i}$, which is the initial ionization rate times its fraction of the initial population. However, as one goes further down the cascade the amplitudes quickly become unwieldy and limiting the number of participating states becomes paramount. For this reason, focus was placed only on the dominant decay channels present in the principle paths seen in Fig.~\ref{fig:2P_path} and Fig.~\ref{fig:2S_path} while weaker channels were ignored.

\begin{figure}[htbp]
\centering
\includegraphics[keepaspectratio, width=0.9\linewidth]{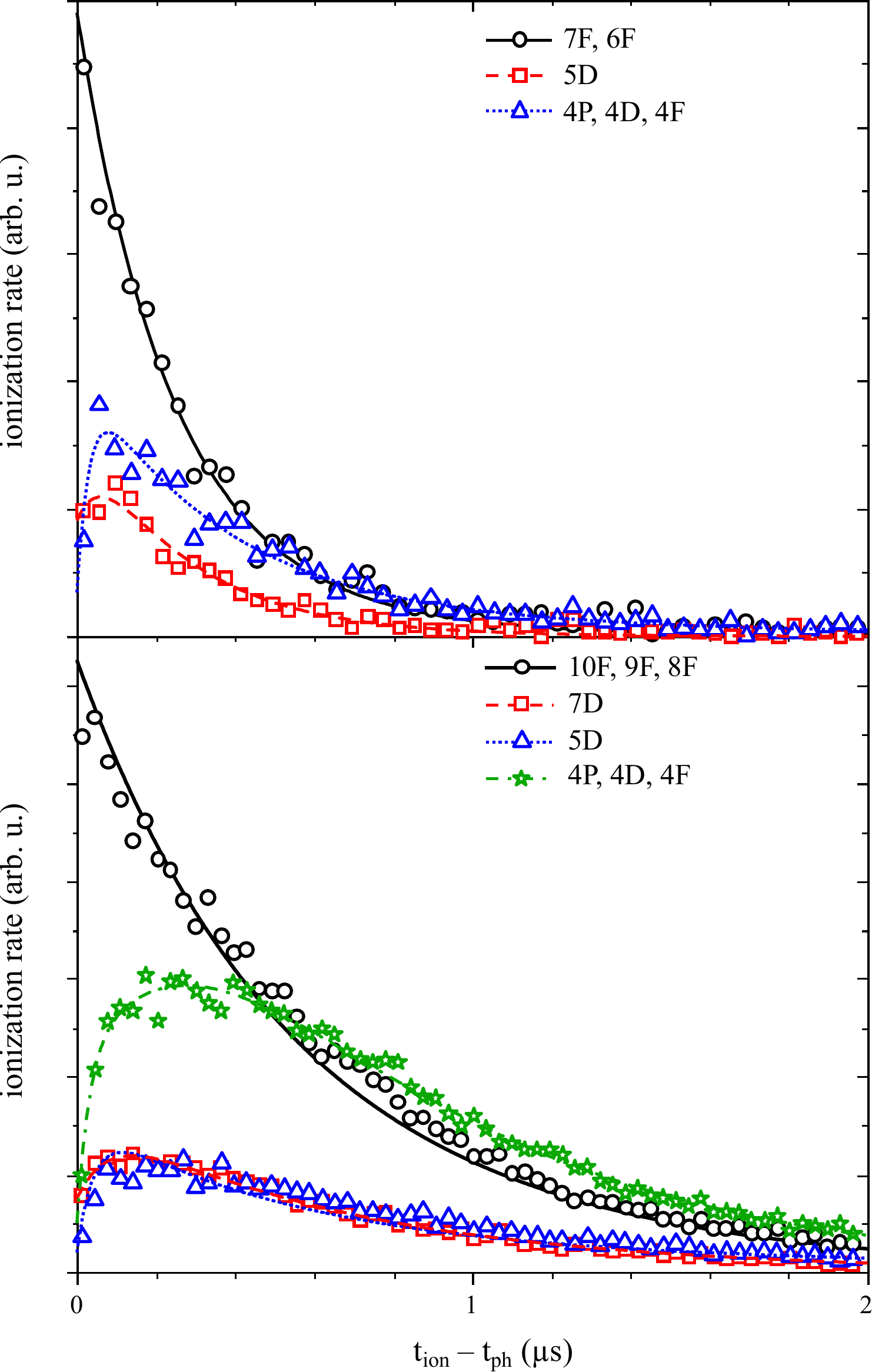}    \caption{Ionization rate versus derived time of ionization for both the 2S (top) and 2P (bottom) sets.}
\label{fig:ionrates_vs_time}
\end{figure}

Fits for each state were made by combining the analytic principle path term, $I_{i}(t)$, and an auxiliary term that characterizes the neglected states by three real-valued fitting parameters $\mathcal{A},\mathcal{B},$ and $\mathcal{C}$:
\begin{equation}
    \mathcal{I}_{\,i}^{\,fit}(t) = \mathcal{A}\:I_{i}(t) + \mathcal{B}\:e^{-\mathcal{C}\,t}
\end{equation}
Applying this ansatz to each state yields the fits seen in Fig.~\ref{fig:ionrates_vs_time}.
Ideally, $\mathcal{A} = 1$ and $\mathcal{B} = 0$ meaning that all participating states were accounted for in the principle path. The fits corresponding to the higher energy levels ($n>4$) matched beautifully with the data and had correspondingly small $\mathcal{B}$ values. However, the lowest energy level has significantly more states neglected in the principle path, which corresponds to a nonneglible $\boldsymbol{B}$. 

For these states, $\mathcal{B}$ represents the sum of the contributing amplitudes from neglected states and decay channels. As can be see in Fig.~\ref{fig:ionrates_vs_time}, this term also corresponds to the initial state populations for the middle states in the cascade that should begin at zero. The coupling parameter $\mathcal{C}$, which in principle is time-dependent (treated as approximately constant here), can be associated with an average over the neglected decay channels initially normalized with respect to $\mathcal{B}$, 
\begin{subequations}
\begin{equation}
    \mathcal{B} = \sum_{p\,\neq\,i,j}A_{p},
\end{equation}
\begin{equation}
        e^{-\mathcal{C}\,t} = \sum_{k\,\neq\,i,j}(\frac{A_{k}}{\mathcal{B}})\,e^{-\Gamma_{k}\,t}.
\end{equation}
\label{eq:AB_fitparams}
\end{subequations}\\
If there was only a single missing dominant channel, then $\mathcal{C}$ roughly corresponds to that channel's coupling constant to the i$^{th}$ state, $\Gamma_{c,i}$. For multiple missing states, then $\mathcal{C}$ changes asymptotically in time from the largest to the smallest coupling constant. Note here that the indices $p$ and $k$ in Eq.~(\ref{eq:AB_fitparams}) are summed over the exact same neglected states.

Each fit was implemented successfully with $\mathcal{A}=1$ and relatively small $\mathcal{B}$ fits except for the lowest $n = 4$ states. For these states, the greatest success was achieved by setting $\mathcal{A}=\mathcal{B}$ and fitting this common parameter. In these final fits the order of magnitude of $\mathcal{B}$ was roughly 50 for the $2P$ set and 1.3 for the $2S$, respectively. This large discrepancy in the $2P$ set primarily originates from the larger amalgam of both initially excited Rydberg and lower lying states that were not accounted for in the principle path. Systems more like the $2S$ set with states that are separated by larger energy gaps are more amenable to this method. 
\section{Conclusion}
In this study, the delayed ionization of $^{6}$Li atoms initially in the $2^{2}S_{1/2}$ and $2^{2}P_{3/2}$ states, due to a narrow bandwidth fs-pulse overlapped with a continuous wave ODT laser, was examined. The addition of a continuous wave laser frustrates the application of the standard ReMi technique as this method relies heavily on knowing when ionization occurs to solve the subsequent equations of motion (i.e. Eq.~(\ref{eq:cyclotron_motion})). Since ionization can now happen at any time between fs-pulses, the time of flight for each fragment is unknown (see also Eq.~(\ref{eq:time_of_flight})).

To overcome this hurdle, coincidence measurements of each ion-electron pair were utilized. The ionization time for each recoil-ion was held fixed while $t_{ion}$ for each electron was varied in order to minimize the total momentum of the pair. This derived time of ionization can then be used to recover the time of flight for each fragment and culminates in the reconstruction of the transverse momentum spectra for the electrons.

These spectra can then be broken down further into the time-independent structure and the time-dependent dynamics. The former can be illustrated via the PAD for each ring in the momentum spectra. Once done, the magnetic dichroism on the states is apparent and the ratio of amplitudes and relative phases for the participating partial waves can be measured. 

The latter is facilitated by the pump-probe scheme consisting of the superposed fs pulse with the continuous ODT laser. Here, the internal population dynamics can be exposed via basic rate equations. The system was modeled as a set of F-orbitals initially excited by the fs-laser that then decay in a cascade back down towards ground. A principle path in the cascade was chosen using the peaks in the photo-electron energy spectrum, the dipole selection rules, and coupling constants that were obtained using the Ark Rydberg Calculator such that only the strongest decay channels remained. Once the path is chosen, a modified ionization rate equation is applied to the delayed ionization and the data fitted accordingly. 

A system with many initially excited Rydberg states (such as the $2P$ set) was found to have a large deviation, seen in the parameter $\mathcal{B}$, as the number of participating states in the cascade was greater than anticipated. However, systems with states that are well separated in energy (seen in the $2S$ set) are better illuminated by this method. Although there is room to improve, such as the optimization method employed to reduce the time resolution, fundamentally the greatest bottleneck lies in the large number of coupled Rydberg states initially excited by the fs-pulse and the associated heavy-computation needed to model the resulting decay. 

The method presented here will be useful in understanding related questions on light-atom interactions that are of interest. For example, in recent experiments employing orbital angular-momentum (OAM) light beams (e.g. \cite{Fang2022}), one must take great care to keep the atoms on the axis of symmetry of the beam so that the angular momentum remains well defined. Atoms are typically too small, relative to the wavelength, to experience the spatial profile of the beam. When off the central axis of the OAM beam the atoms are effectively in a plane wave. However, Rydberg atoms such as those explored here, are significantly larger and have wave functions that can easily extend out to a hundred or more atomic units. At this scale, the spatial profile of the beam is available to the atom and experiments that rely the violation of the dipole approximation can be readily investigated.  

\section*{Acknowledgments}
The experimental material presented here is based upon work supported by the National Science Foundation
under Grant \hbox{No.~PHY-1554776} and \hbox{No.~PHY-2207854}.
\bibliography{Rydberg}

\end{document}